\documentclass[12pt]{elsarticle}
\usepackage{graphicx}
\usepackage{subfigure}
\usepackage{epstopdf}
\usepackage[refpage]{nomencl}
\usepackage{caption}
\usepackage{float}
\usepackage{array}
\usepackage{amsmath}
\usepackage{amsfonts}
\usepackage{pdfpages}
\biboptions{sort&compress}

\journal{Phys. Lett. A}

\begin{document}

\begin{frontmatter}

\title{Absence of long-range order in a three-dimensional stacked Ising antiferromagnet on kagome lattice}

\author{M. Semjan}

\author{M. \v{Z}ukovi\v{c}\corref{cor}}
\ead{milan.zukovic@upjs.sk}

\address{%
Department of Theoretical Physics and Astrophysics, Institute of Physics, Faculty of Science, Pavol Jozef \v{S}af\'arik University in Ko\v{s}ice, Park Angelinum 9, 041 54 Ko\v{s}ice, Slovak Republic
}%

\date{\today}

\cortext[cor]{Corresponding author}


\begin{abstract}
  
We study the possibility of long-range ordering (LRO) in a 3D system of vertically stacked layers of Ising antiferromagnet on a kagome lattice (SIAKL). Monte Carlo simulations are carried out for a varying interlayer coupling strength and a finite-size scaling analysis is performed for selected cases. Unlike in the related Ising system on a triangular lattice, in which even a finite number of the layers can stabilize LRO, or the Heisenberg system on the same kagome lattice, in which LRO emerged in 3D for a sufficient strength of the interlayer coupling, no LRO could be observed in the present model. This makes SIAKL a rare example of a 3D Ising paramagnet composed of frustrated layers coupled by unfrustrated interaction.

\end{abstract}


\end{frontmatter}

\section{Introduction}

It is well known that frustration in general disturbs formation of long-range order (LRO) in spin systems. This is particularly the case in two-dimensional (2D) systems, with the prototypical examples being the geometrically frustrated Ising antiferromagnets on a triangular~\cite{wannier50} (IATL) and a kagome~\cite{syozi1951,kano1953} (IAKL) lattice, which show no LRO at any finite temperature. Nevertheless, such systems with infinitely degenerate ground states are very susceptible to different perturbations, which can partially relieve the frustration and can lead to some type of LRO driven by thermal fluctuations. 

For example, the increased magnitude of the spin $S$ in the 2D IATL model was found to lead to a so-called partial-disordered (PD) structure $(M,0,-M)$ with two sublattices ordered antiferromagnetically and the third one disordered~\cite{netz1993,nagai1993,yama1995,lipo1995}. Similar ordering was also confirmed in the actively studied 3D stacked IATL (SIATL) model~\cite{berker84,blankschtein84,coppersmith85,heinonen89,netz91a,netz91b,bunker93,plumer93,plumer94,plumer95,kim90,nagai94,kurata95,todoroki2004,meloche07,zukovic2012,lin2014,liu2016,borovsky2016,zukovic2018} due to additional unfrustrated interactions in the stacking direction along the $z$ axis. In particular, the system has been found to undergo the phase transition from the paramagnetic to the PD phase, which belongs to the 3D XY universality class~\cite{berker84,blankschtein84,plumer93,bunker93,meloche07}, albeit the tricritical behavior has also been suggested~\cite{heinonen89}. However, the PD phase does not persist to very low temperatures. Some studies suggested another phase transition occurring at lower temperatures a to a ferrimagnetic phase $(M,-M/2,-M/2)$, with one sublattice fully ordered and two partially disordered~\cite{blankschtein84,netz91a,todoroki2004}. On the other hand, several other studies argued that due to linear-chain-like excitations in the system there is a crossover to the low-temperature phase, which is a 3D analog of the 2D Wannier phase~\cite{coppersmith85,heinonen89,kim90,borovsky2016,zukovic2018}. In the IATL to SIATL dimensional crossover study it was found that the PD phase can also develop in the system consisting of a sufficiently large but finite number of the IATL layers stacked along the $z$ axis. Moreover, it is preceded by another quasi-LRO phase with peculiar pseudocritical correlations of the Berezinskii-Kosterlitz-Thouless (BKT) type~\cite{lin2014}. 

Stacking frustrated lattices leads to LRO also in some other 3D systems, such as the stacked Villain lattice (frustration caused by competing interactions)~\cite{nagai85} or the face-centered cubic~\cite{liebmann86} and hexagonal-close-packed lattices with antiferromagnetic nearest-neighbor interaction (frustration caused by lattice structure)~\cite{auerbach88,hoang12}. Nevertheless, there are examples of 3D Ising spin systems, which show paramagnetic behavior at all finite temperatures. One possibility is to artificially design such systems from the IATL or the Villain lattices by using a glue-Hamiltonian~\cite{horiguchi87} but there are also some realistic systems, such as a chessboard (CB3) lattice~\cite{diep2013} or a pyrochlore lattice~\cite{anderson56,liebmann86}. The latter is composed of alternating IATL and IAKL planes, forming corner-sharing tetrahedra, with the frustration both within and between the planes. Frustrated stackings can lead to a classical spin-liquid regime at low temperature with strong correlations but no LRO also in SIATL if the interlayer coupling is much weaker than the intraplane one~\cite{liu2016}.

The IAKL system can be constructed from the IATL model by periodical removal of a quarter of sites, 
which results in even greater ground state degeneracy and no LRO at any, including zero,
temperature~\cite{syozi1951, kano1953}. In the ground state, unlike IATL with the algebraically decaying
correlation function, IAKL displays only short-range ordering with two spins aligned parallel and 
the third one antiparallel in each elementary triangular plaquette. The question is: can unfrustrated
stacking of such IAKL planes (3D SIAKL model) lead to some kind of LRO, as it was in the SIATL case, or the system will remain paramagnetic at all temperatures? On one hand, IAKL is a highly frustrated and degenerate lattice of corner-sharing triangles and therefore it can be viewed as a natural 2D analog of the pyrochlore lattice of corner-sharing tetrahedra, which shows no LRO. On the other hand, the pyrochlore lattice is frustrated both within the planes as well as in the stacking direction, while SIAKL is only frustrated within the planes.

To our best knowledge, the effect of the interlayer coupling has only been studied in the case of the Heisenberg model.
In particular, the Heisenberg antiferromagnet on the kagome lattice (HAKL) with the spin $S=1/2$ received
the most attention and the general consensus is that in 2D there is no LRO, just like in IAKL~\cite{muller2018} 
(although, in the Heisenberg case, the reason is the Mermin–Wagner theorem~\cite{mermin1966}). 
The possibility of the emergence of magnetic ordering in ground state by stacking of 2D HAKL layers 
(SHAKL model) was examined~\cite{schmalfuss04,gotze16}. The initial results, obtained by using a spin-rotation-invariant Green's function method and the linear spin wave theory~\cite{schmalfuss04}, showed that the additional interlayer coupling 
of any sign and strength was unable to stabilize LRO. However, the consequent study~\cite{gotze16}, which used a more precise
coupled-cluster method to higher orders of the approximation claimed that if the strength of the interlayer 
coupling (almost independent of its sign) exceeds about 15\% of the intralayer one, then the LRO can be established. 

In the present paper we investigate the behavior of the SIAKL system, consisting of IAKL layers vertically stacked along the $z$ axis and coupled by (unfrustrated) interlayer interaction of varying strength. The goal it to answer the question regarding the existence of any kind of (quasi-)LRO at finite temperatures.

\section{Model and method}

The total Hamiltonian of the studied SIAKL system can be considered to be composed of the intralayer 
and interlayer contributions, which are given by the following equations:

\begin{equation}
  \begin{aligned}
    \mathcal{H}_{xy} &= -J_1\sum_{\langle i,j\rangle }{\sigma_i\sigma_j}, \\ 
    \mathcal{H}_{z}  &= -J_2\sum_{\langle i,k\rangle }{\sigma_i\sigma_k}, \\
    \mathcal{H}      &= \mathcal{H}_{xy} + \mathcal{H}_{z},
  \end{aligned}
  \label{eq:H}
\end{equation}
where $\mathcal{H}_{xy}$ represents the energy within the $xy$ layers, $\mathcal{H}_{z}$ is the interlayer energy (or
energy of chains along $z$ axis) and $\mathcal{H}$ gives the total energy of the system. The summations $\langle i,j\rangle$  and $\langle i,k\rangle$ go over all nearest neighbors within the planes and within the chains along the $z$ axis, respectively. $\sigma_i = \pm 1$ are the Ising variables and $J_1$ and $J_2$ are the coupling constants inside the layers and between the layers, respectively. $J_1$ is chosen to be $J_1=-1$ (antiferromagnetic) to introduce frustration within the layers, while $J_2>0$ is for convenience considered ferromagnetic\footnote{In the absence of the symmetry-breaking field both choices $J_2>0$ and $J_2<0$ would mean that the spins along $z$ axis remain unfrustrated, leading to the same critical behavior.}.

We assume that the SIAKL lattice consists of three interpenetrating sublattices. The sublattice
magnetizations can be calculated as
\begin{equation}
  M_{\alpha} = \sum_{i\in \alpha}\sigma_i,\qquad \alpha = 1,~2,~3,
\end{equation}
and the total magnetization $M$ is then obtained as $M=M_1+M_2+M_3$. 

Several other derived quantities are calculated using the following equations (we set the Boltzmann constant $k_B=1$): the internal energy per spin
\begin{equation}
  e = \frac{\langle \mathcal{H} \rangle}{N},
  \label{eq:e}
\end{equation}
the specific heat
\begin{equation}
  C = \frac{1}{NT^2}\big(\langle \mathcal{H}^2 \rangle - \langle \mathcal{H} \rangle^2\big),
  \label{eq:c}
\end{equation}
the specific heat of the IAKL layers
\begin{equation}
  C_{xy} = \frac{1}{NT^2}\big(\langle \mathcal{H}_{xy}^2 \rangle - \langle \mathcal{H}_{xy} \rangle^2\big),
  \label{eq:cxy}
\end{equation}
and the specific heat of the chains in the $z$ axis direction
\begin{equation}
  C_{z} = \frac{1}{NT^2}\big(\langle \mathcal{H}_{z}^2 \rangle - \langle \mathcal{H}_{z} \rangle^2\big).
  \label{eq:cz}
\end{equation}
The mean values of the sublattice magnetizations are obtained as
\begin{equation}
  m_{\alpha} = \frac{\langle M_{\alpha} \rangle}{N},
  \label{eq:mi}
\end{equation}
and the sublattice magnetic susceptibilities as
\begin{equation}
  \chi_{\alpha} = \frac{1}{NT}\big(\langle M_{\alpha}^2 \rangle - \langle M_{\alpha} \rangle^2\big),
  \label{eq:chii}
\end{equation}
where $T$ is the temperature, $N$ is the total number of spins, and $\alpha = 1,~2,~3$. To break the symmetry and to prevent the sublattice magnetizations from flipping between the three sublattices, at each MC sweep we consider $M_1 \equiv M_{max}=\max\{M_1,M_2,M_3\}$, $M_{2} \equiv M_{median}={\rm med}\{M_1,M_2,M_3\}$, and $M_3 \equiv M_{min}=\min\{M_1,M_2,M_3\}$.

Furthermore, we define the quantity $o_{z}$, which describes ordering of the chains along the $z$ axis
\begin{equation}
  o_{z} = \frac{1}{N}\Bigg\langle\sum_{Chains}\Bigg|\sum_{i=1}^{L}\sigma_i\Bigg|\Bigg\rangle,
  \label{eq:oz}
\end{equation}
where the outer summation goes through all the chains, the inner one over all spins in the given chain, 
and $L$ is the chain length (number of the IAKL layers).

We also calculate the entropy density (entropy per spin), using the thermodynamic integration method (TIM)~\cite{kirkpatrick1997}: 
\begin{equation}
  s(\beta) = \ln{2} + \beta e(\beta) - \int_0^\beta e(\beta')d\beta',
  \label{eq:tim}
\end{equation}
where the value in the argument of the natural logarithm is the multiplicity of the local degrees of freedom and $\beta=1/T$
is the inverse temperature.

The model is studied by employing Monte Carlo (MC) simulations with the standard single spin-flip Metropolis algorithm implemented on the highly parallel architecture of GPU. We execute extensive MC simulation runs on the 3D SIAKL lattices with the linear sizes ranging from $L = 16\ (N=12\ 288)$ up to $L=128\ (N=6\ 291\ 456)$, and apply the periodic boundary conditions\footnote{Some preliminary MC simulations were performed for systems with a finite number of layers and open boundary conditions in the stacking direction but finding no traces of LRO we fully focused on the 3D model emulated by the cubic $L \times L \times L$ shape and the periodic boundary conditions.}. The simulations start from random states at sufficiently high temperatures and continue to lower temperatures starting from the last configuration obtained at the previous step. For the entropy calculation it is convenient to consider the inverse temperature, which starts at $\beta=0$ ($T=\infty$) and increases with a non-uniform step up to $\beta=794$ ($T\approx 0.0012$). The total number of temperature steps is $N_T=100$. The density of the temperature mesh is selected as to minimize the error in the numerical calculation of the integral in TIM. The total number of MC sweeps is $1.5\times 10^6$ with the initial $5\times 10^5$ sweeps discarded for equilibration.

We would like to remark, that in general for frustrated spin systems it is advisable to use some more sophisticated MC techniques, such as the replica exchange (RE) method~\cite{hukushima96}, to prevent potential problems (especially at low temperatures) with reaching equilibrium due to long relaxation times or getting trapped in metastable states. Such problem may also potentially occur in the IAKL model, for example in the presence of magnetic impurities and external field, in which case the RE method might be desirable~\cite{soldatov19}. Fortunately, in the present model the knowledge of the ground-state energy allows verification whether or not the true equilibrium has been reached and, as the results presented below indicate, such problems do not occur here even for the largest lattice sizes. Even the spin ``freezing'' phenomenon, which is observed in the present model, is not as severe as in SIATL~\cite{netz91a,borovsky2016,zukovic2018} and does not seem to prevent the system from reaching the ground state. On the other hand, applying the RE method to simulations of larger system sizes comprising millions of spins (which due to strong finite-size effects are needed for a reliable finite-size scaling analysis) would require enormous memory demands, since a large number of replicas must be kept simultaneously in the memory throughout the simulations.

\section{Results and discussion}

\subsection{Effect of varying interlayer interaction}

\begin{figure}[t!]
  \center
  \subfigure{\includegraphics[scale=0.45,clip]{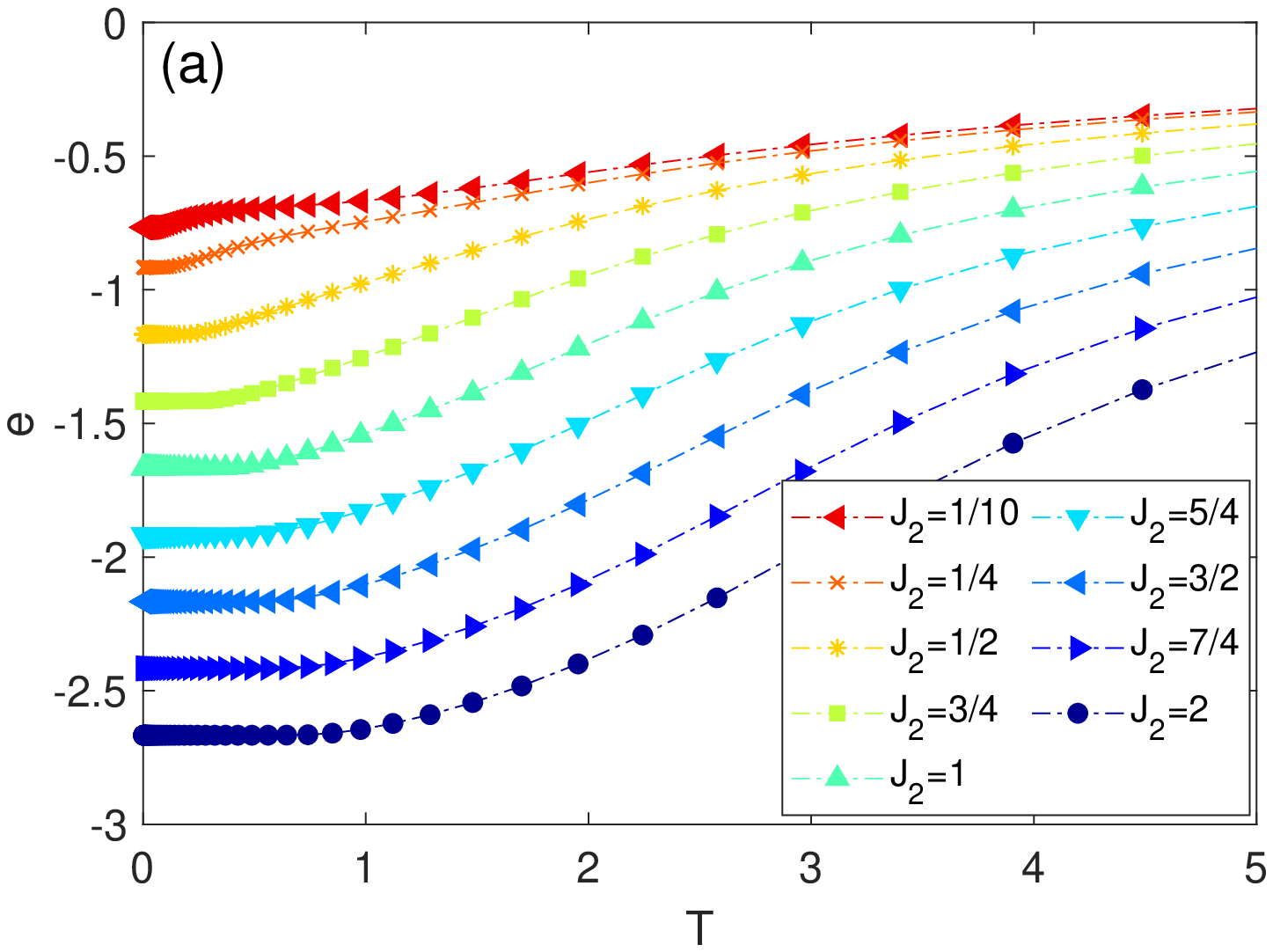}\label{fig:ene}}
  \subfigure{\includegraphics[scale=0.45,clip]{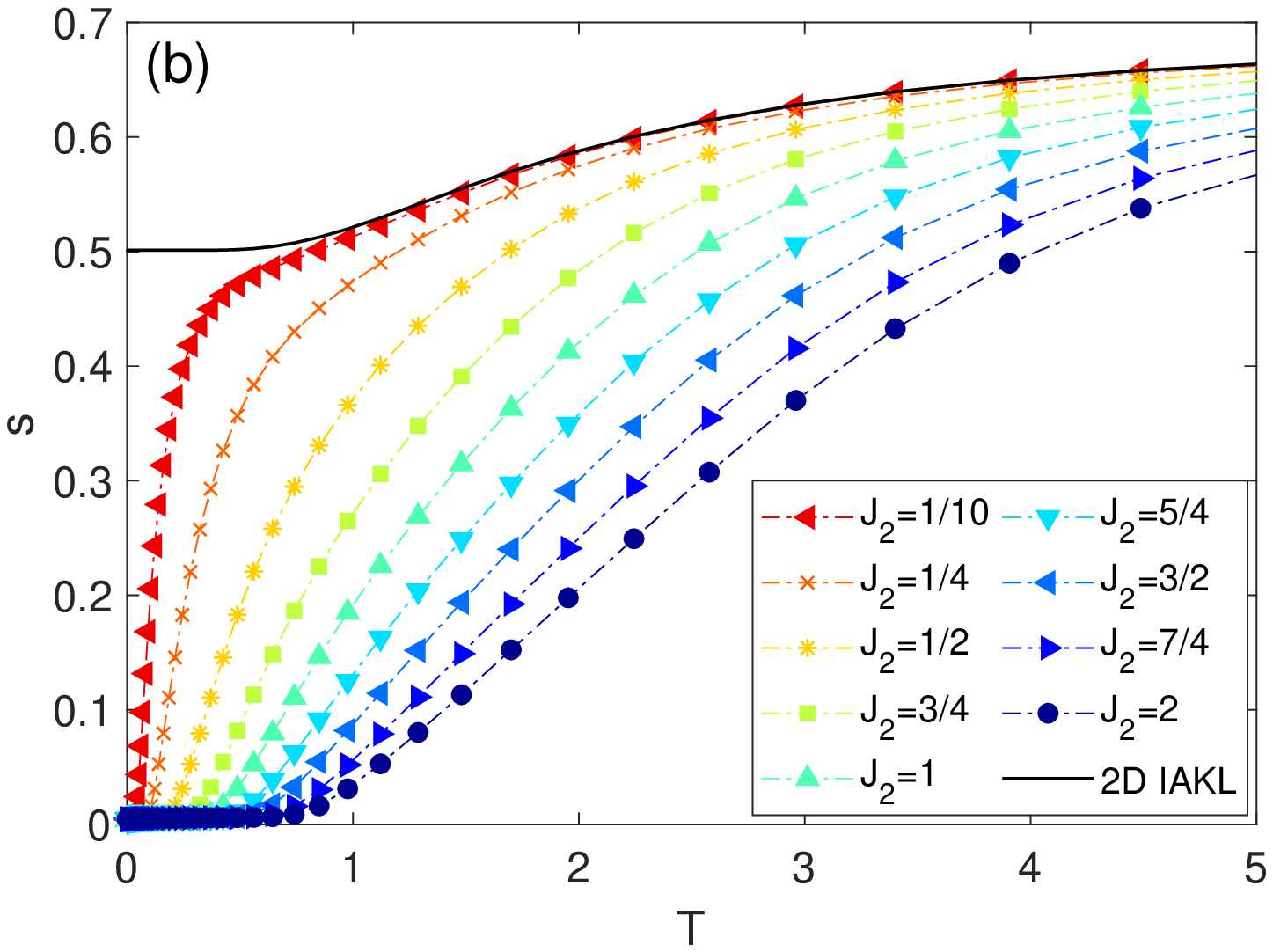}\label{fig:ent}}
  \caption{(a) The internal energy $e$ and (b) the entropy density $s$, for different values of $J_2$ and $L=64$.}
  \label{fig:e}
\end{figure}

Below we present temperature dependencies of the calculated quantities for a varying interlayer
coupling constant, $J_2=1/10$, $1/4$, $1/2$, $3/4$, $5/4$, $3/2$, $7/4$ and $2$, 
for the linear lattice size $L=64$. The internal energy curves shown in Fig.~\ref{fig:ene} display 
smooth variations with the temperature, terminating at the lowest temperatures close to $T=0$ with
the values expected in the ground state (GS). Namely, at $T=0$ the minimum energy is realized for the states
with the neighboring spins on each triangular plaquette within the planes arranged ferrimagnetically ($\uparrow\uparrow\downarrow$ or $\uparrow\downarrow\downarrow$) and the neighboring spins in the $z$-axis
direction aligned ferromagnetically. Such an arrangement will result in the GS energy
\begin{equation}
  e_{GS} = -\frac{2}{3}|J_1| - J_2,
  \label{eq:egs}
\end{equation}
which is in good agreement with the lowest-temperature values from the MC simulations. 
The differences are of the order of $10^{-7}$ and are most likely caused by rounding errors.

The behavior of the entropy density curves is shown in Fig.~\ref{fig:ent}. For illustration, we also included the curve corresponding to $J_2=0$, i.e., a single IAKL plane. It is well known that the latter terminates at $T=0$ 
showing a very large residual value of $s_0=0.5018$~\cite{kano1953}. Nevertheless, once the interlayer interaction $J_2$ is turned on, instead of leveling off at some finite values, the entropy curves on approach to $T=0$ take a sudden plunge towards a value close to zero. In particular, the estimated value of the residual entropy for all the curves with $J_2 > 0$ and $L=64$ is $s_o(L=64) \approx 0.003$. 

\begin{figure}[t!]
  \center
  \subfigure{\includegraphics[scale=0.45,clip]{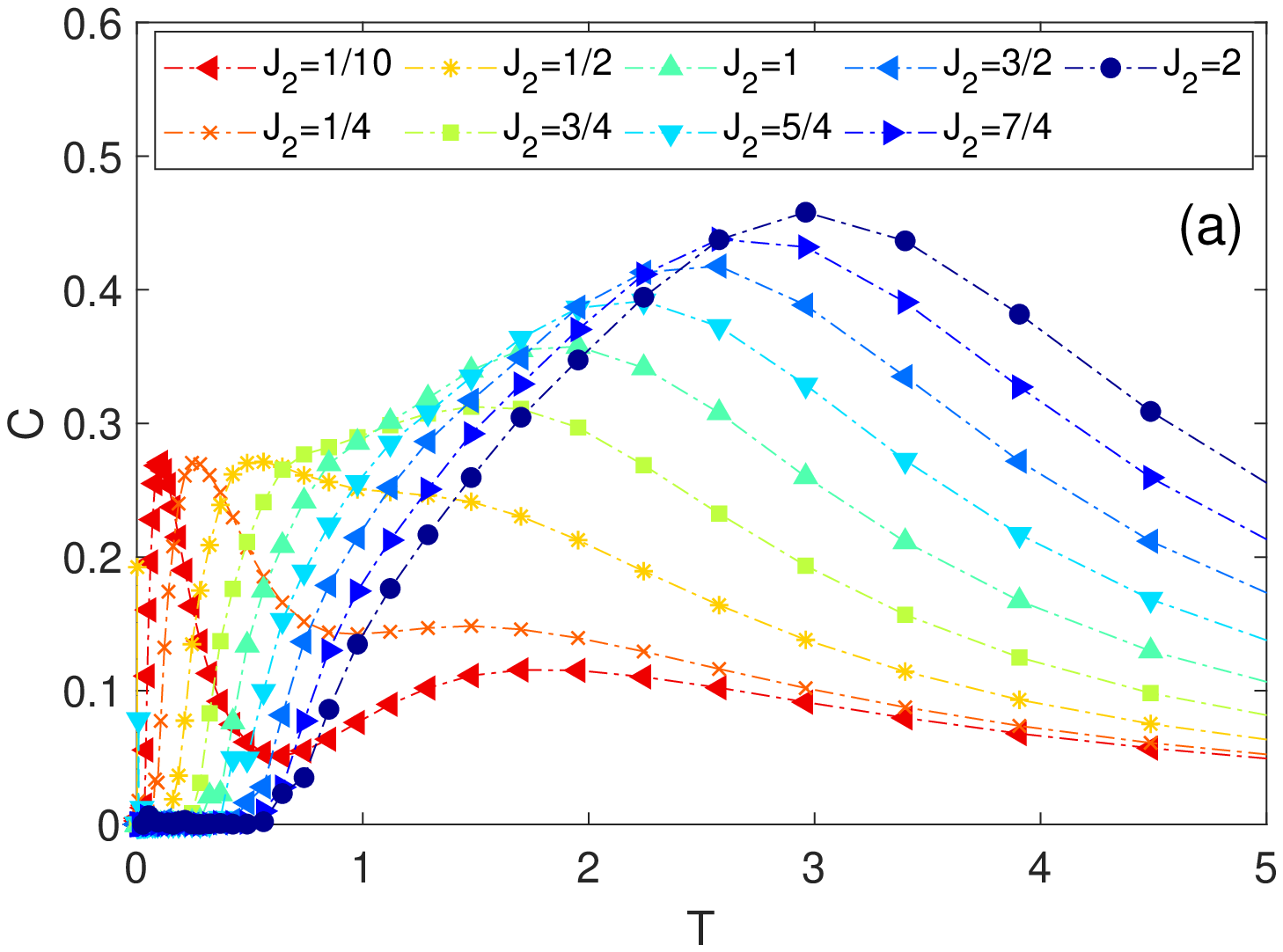}\label{fig:all_J2}}
  \subfigure{\includegraphics[scale=0.45,clip]{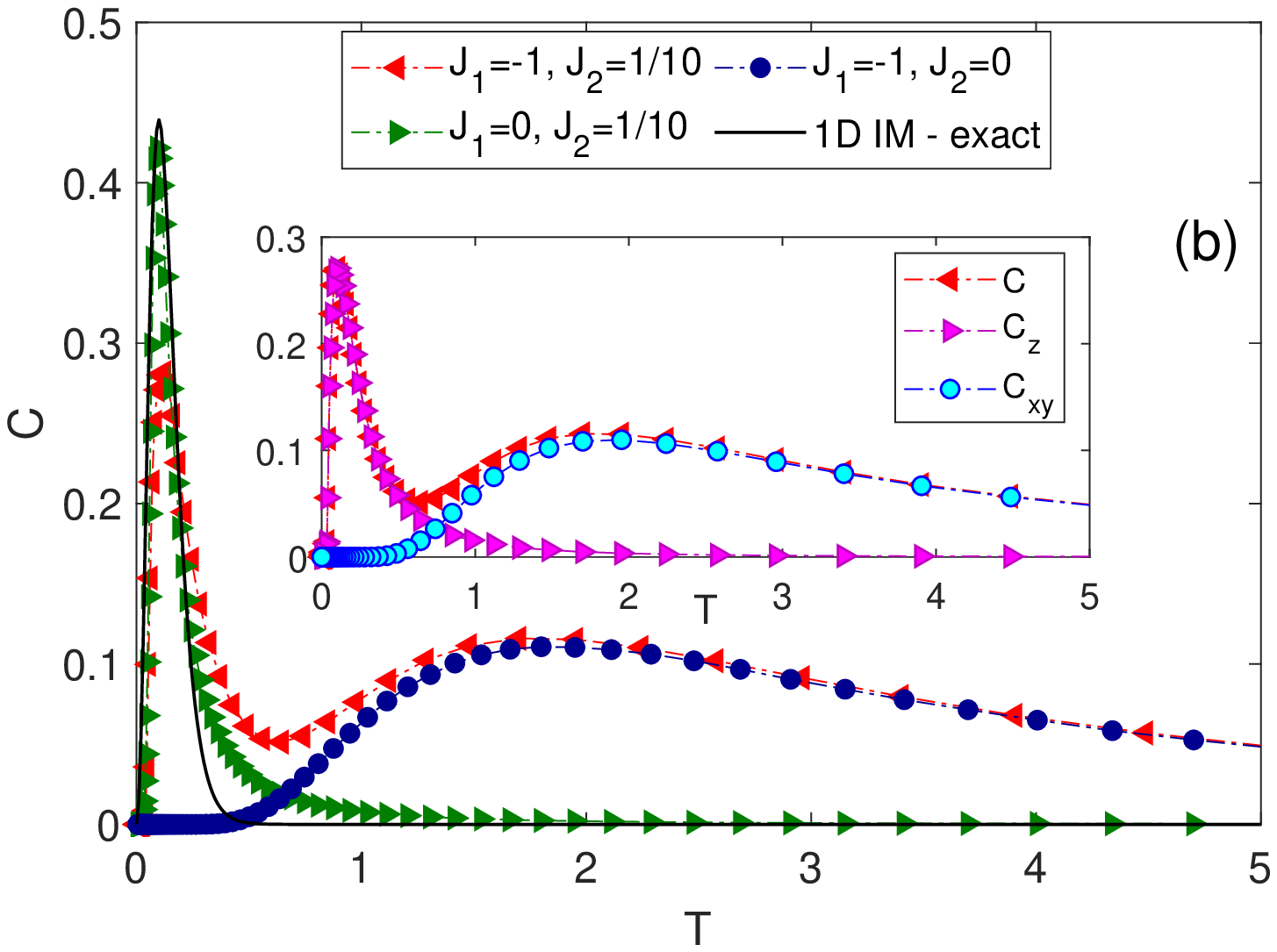}\label{fig:lim_cases}}
  \caption{(a) The specific heat as a function of temperature, for different values of $J_2$. (b) The specific heat curves for $J_2=1/10$ (red left-pointing triangles) and the two limiting cases of $J_1=0$, i.e., 1D ferromagnetic chains (green right-pointing triangles) and $J_2=0$, i.e., 2D IAKL (blue filled circles). For the former also the exact solution is plotted (black solid curve). The inset shows the $C$ curve for $J_2=1/10$ with the axial ($C_z$) and in-plane ($C_{xy}$) components.}
  \label{fig:c}
\end{figure}

In the temperature dependencies of the corresponding specific heat curves we observed the formation of two peaks for smaller $J_2$, which merged to one at larger $J_2$ (Fig.~\ref{fig:all_J2}). With the increasing value of $J_2$ the two peaks evolve as follows. The height of the sharper low-temperature peak remains unchanged but its width becomes gradually larger and its position shifts to higher temperatures. On the other hand, the height of the broader high-temperature peak gradually increases without a noticeable change of its position. This behavior makes the two peaks merge above $J_2=1/2$ into one broad peak with a shoulder-like shape from the low-temperature side. With further increase of $J_2$ the resulting single peak continues shifting to higher temperatures and becomes both broader and higher.

We believe that the observed anomalies in the specific heat can be attributed to non-critical spin excitations in the IAKL planes and within the chains along the $z$ axis. To clarify the origin of the two peaks let us consider the case of $J_2=1/10$, for which they are the most distinctly separated. In Fig.~\ref{fig:lim_cases} we superimpose in the same plot the curves for the 1D Ising chain and the 2D IAKL model, which are the limiting cases of the present SIAKL model for $J_1=0$ and $J_2=0$, respectively, and are known to show no LRO at any finite temperature. Since the former case has also an exact solution ($C=T^2/[J_2^2\cosh^2(T/J_2)])$, we also include it. Apparently, the position of the low-temperature peak of the SIAKL model with $J_2=1/10$ coincides very well with the peak of the 1D model and the position of the high-temperature peak with that of the 2D IAKL model. Also the inset, showing separately the specific heat contributions $C_z$ and $C_{xy}$, confirms that the former corresponds to the excitations along the chains direction and the latter to the intraplane excitations.

%
%

\begin{figure}[t!]
  \center
  \subfigure{\includegraphics[scale=0.45,clip]{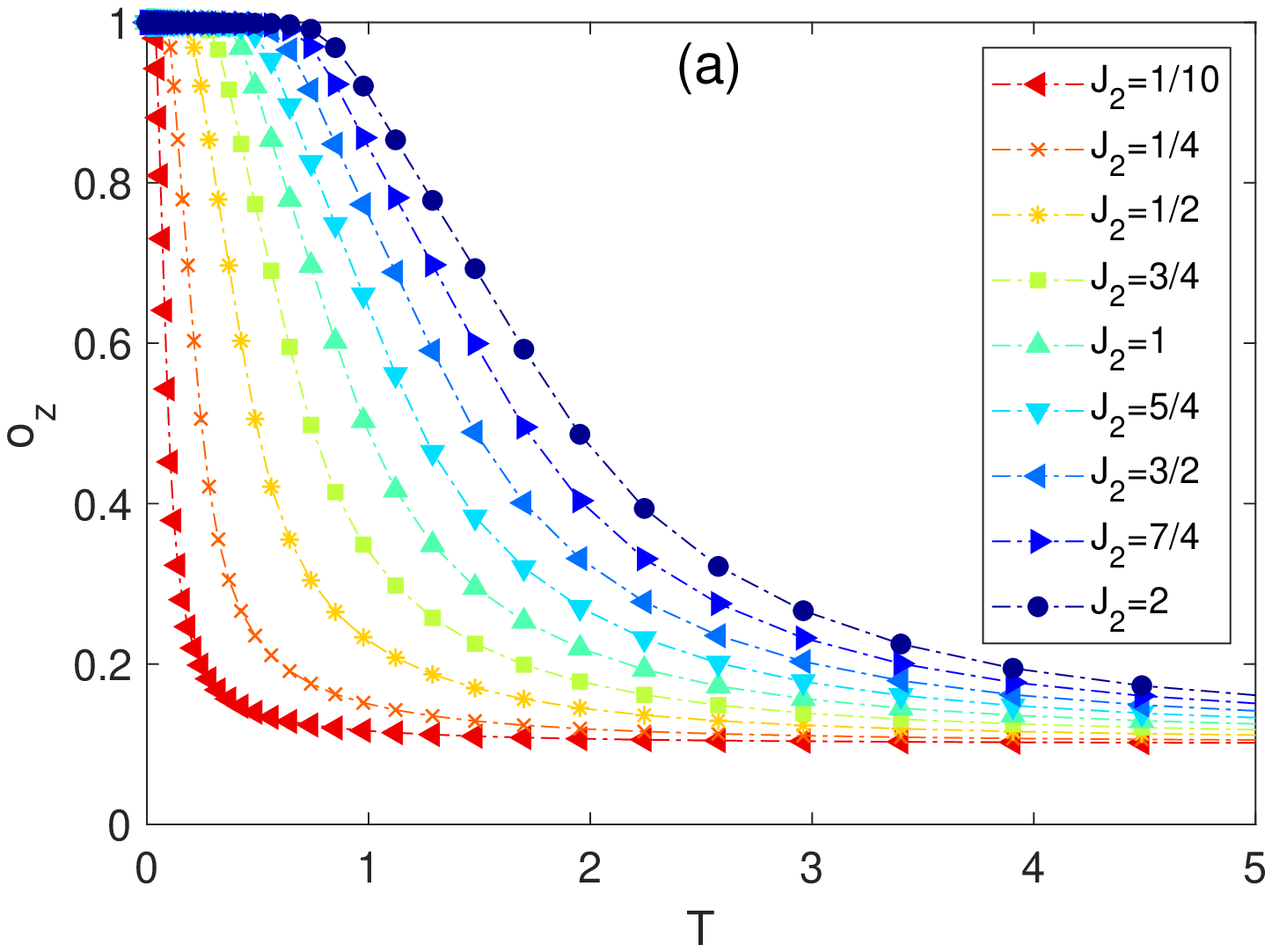}\label{fig:oz_a}}
  \subfigure{\includegraphics[scale=0.45,clip]{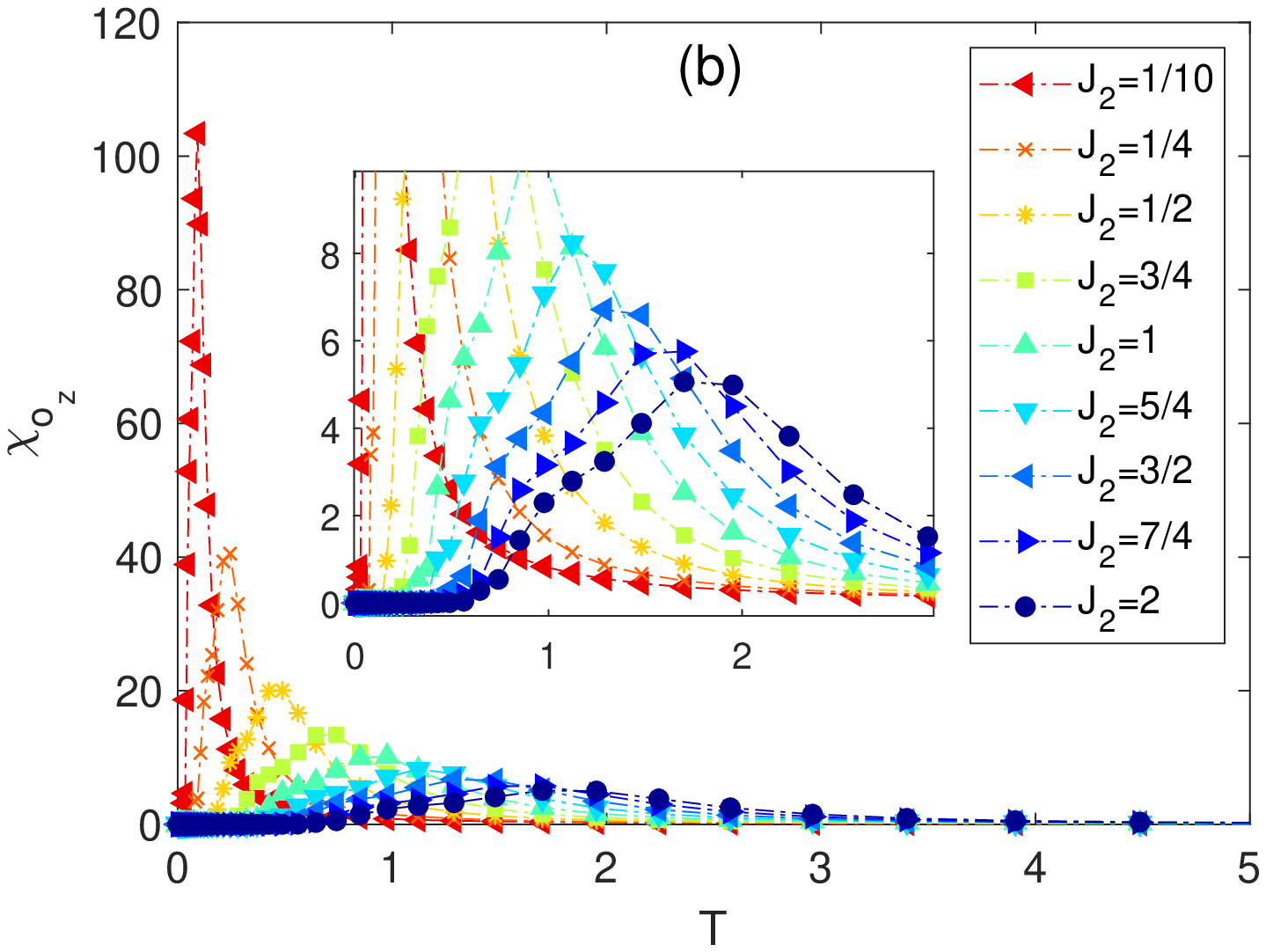}\label{fig:chiz_a}}
  \caption{(a) The parameter of spin ordering along $z$ axis, $o_z$, for different values of $J_2$, and (b) its susceptibility $\chi_{o_z}$. The inset in (b) focuses on low and broad peaks for larger $J_2$.}
  \label{fig:oz}
\end{figure}

Furthermore, in Fig.~\ref{fig:oz} we present the temperature dependence of the chain order parameter $o_z$ and it's susceptibility $\chi_{o_z}$. They demonstrate that the onset of the chain ordering corresponds very well with the low-temperature peak in the specific heat. Below this temperature, the spins align ferromagnetically along the $z$-axis, and since there is no frustration, $o_z$ approaches the saturation value 1 at sufficiently low temperatures. 

\subsection{Finite-size effects}

For finite system sizes various anomalies in physical quantities are smeared and rounded, even singularities related to phase transitions. The latter can be reliably detected by performing a finite-size scaling (FSS) analysis. 
For that purpose we selected two cases of a small and large interlayer couplings, with $J_2=1/10$ and $J_2=2$, and studied their FSS behavior for different lattice sizes $L = 16$ ($12\ 288$ spins in total), $32$ ($98\ 304$ spins), $64$ ($786\ 432$ spins), $96$ ($2\ 654\ 208$ spins), and $128$ ($6\ 291\ 456$ spins).  

\subsubsection{Small interlayer coupling}

\begin{figure}[t!]
  \center
  \subfigure{\includegraphics[scale=0.45,clip]{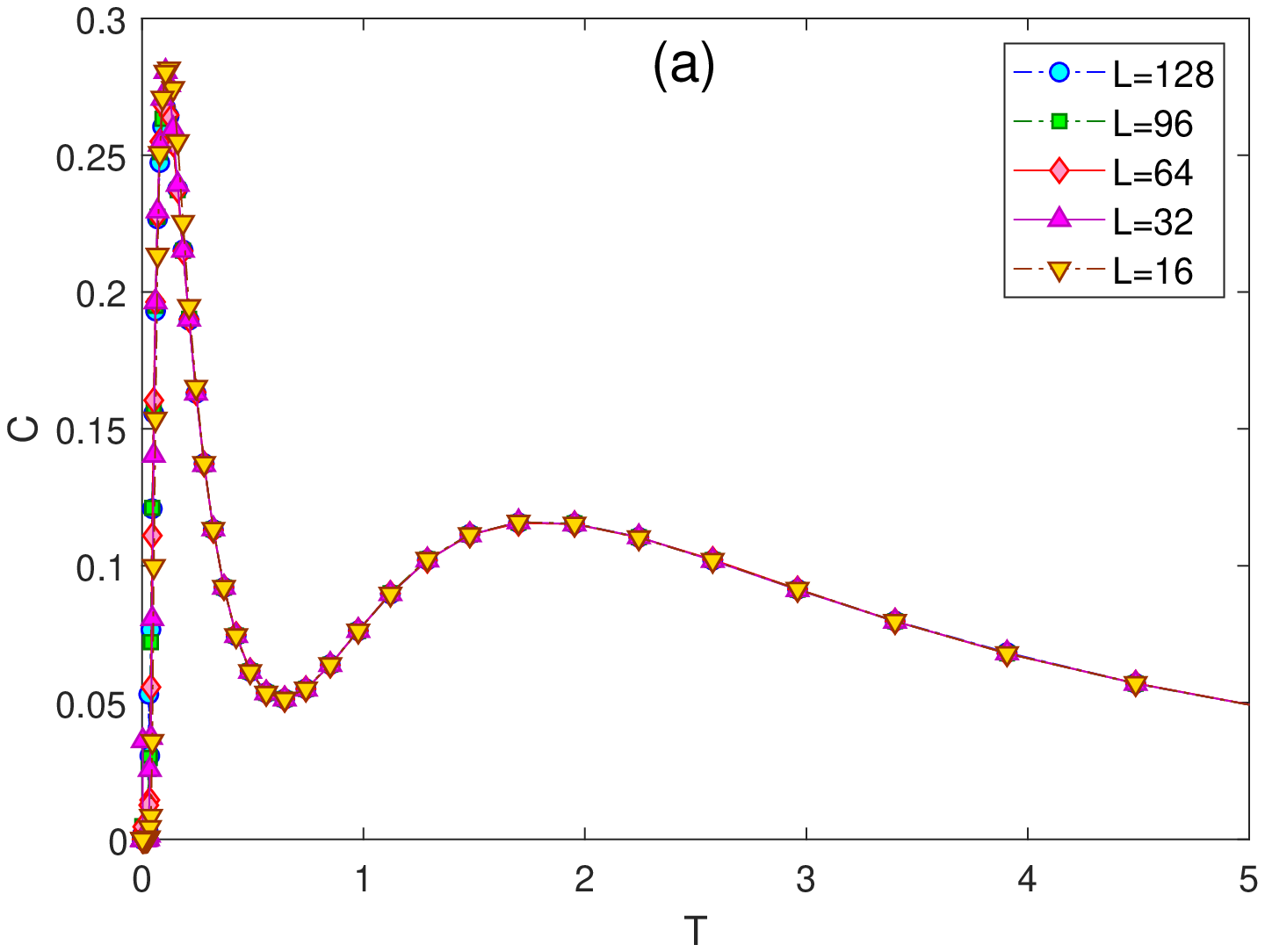}\label{fig:c_fixedJ2}}
  \subfigure{\includegraphics[scale=0.45,clip]{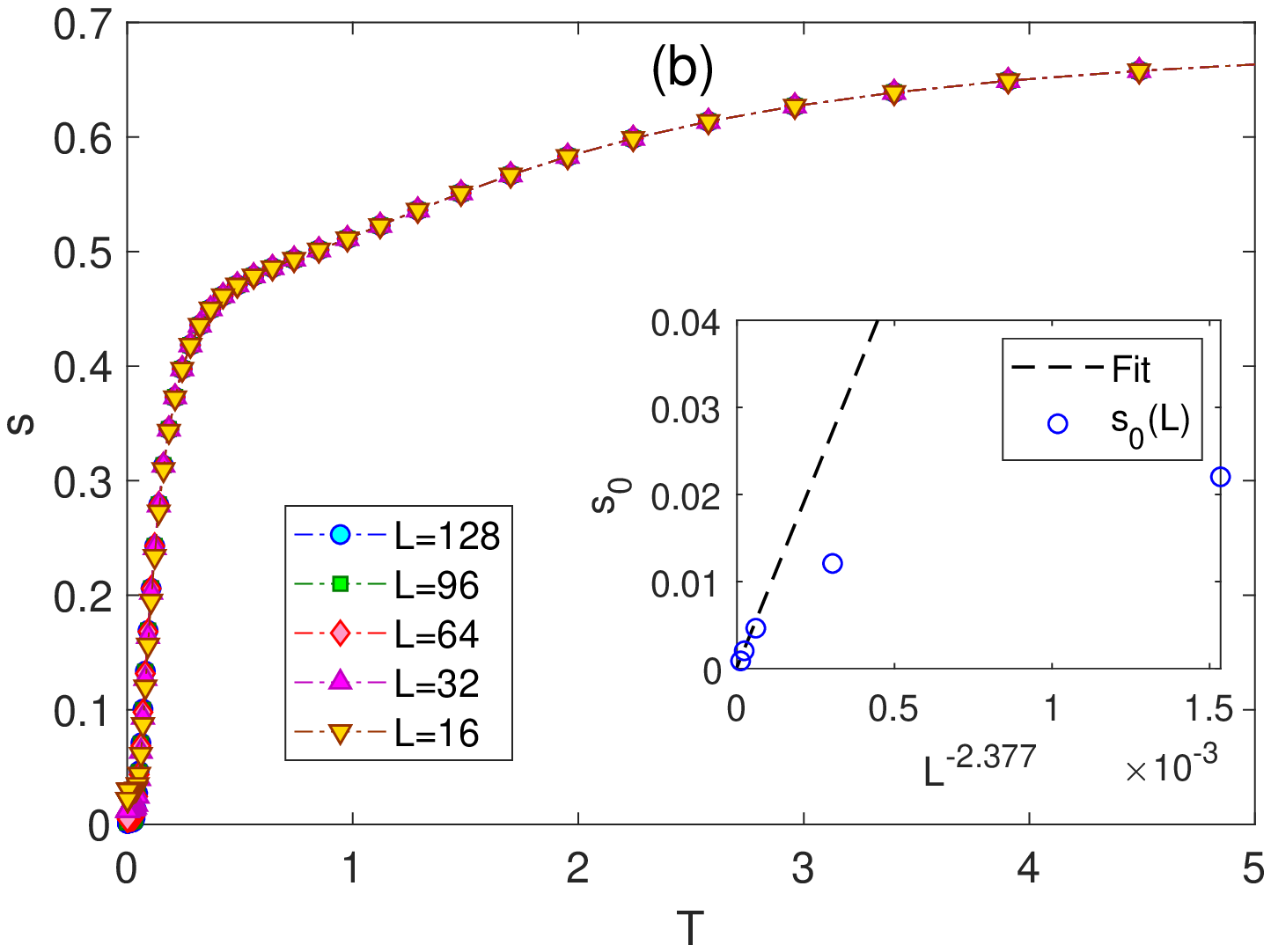}\label{fig:s_fixedJ2}}
  \caption{(a) The specific heat $C$ and (b) the entropy density $s$, for different values of $L$ and $J_2=1/10$. The inset in (b) shows the extrapolation of the residual entropy values to thermodynamic limit $s_o(L \to \infty)=0.00013 \pm 0.00295$, considering only the largest sizes $L$ complying with the asymptotic linear regime.}
  \label{fig:cfss}
\end{figure}

Let us first focus on the $J_2=1/10$ case, which shows two distinct anomalies in the specific heat. As shown in Fig.~\ref{fig:c_fixedJ2}, both peaks are rather insensitive to the lattice size and show no systematic variation with $L$. This behavior corroborates our above assumption of no phase transition and thus no LRO at any finite temperatures. Nevertheless, we can observe some finite-size variation in the entropy density, presented in Fig.~\ref{fig:s_fixedJ2}. In particular, the lowest-temperatures values, which approximate the residual (zero-point) entropy, show a systematic decrease with $L$. The thermodynamic limit value, obtained by considering the power-law decay and excluding the smaller lattice sizes that do not comply with the asymptotic linear regime on a log-log scale (see the inset of Fig.~\ref{fig:s_fixedJ2}), is estimated as $s_o(L \to \infty)=0.00013 \pm 0.00295$.

\begin{figure}[t!]
  \center
  \subfigure{\includegraphics[scale=0.45,clip]{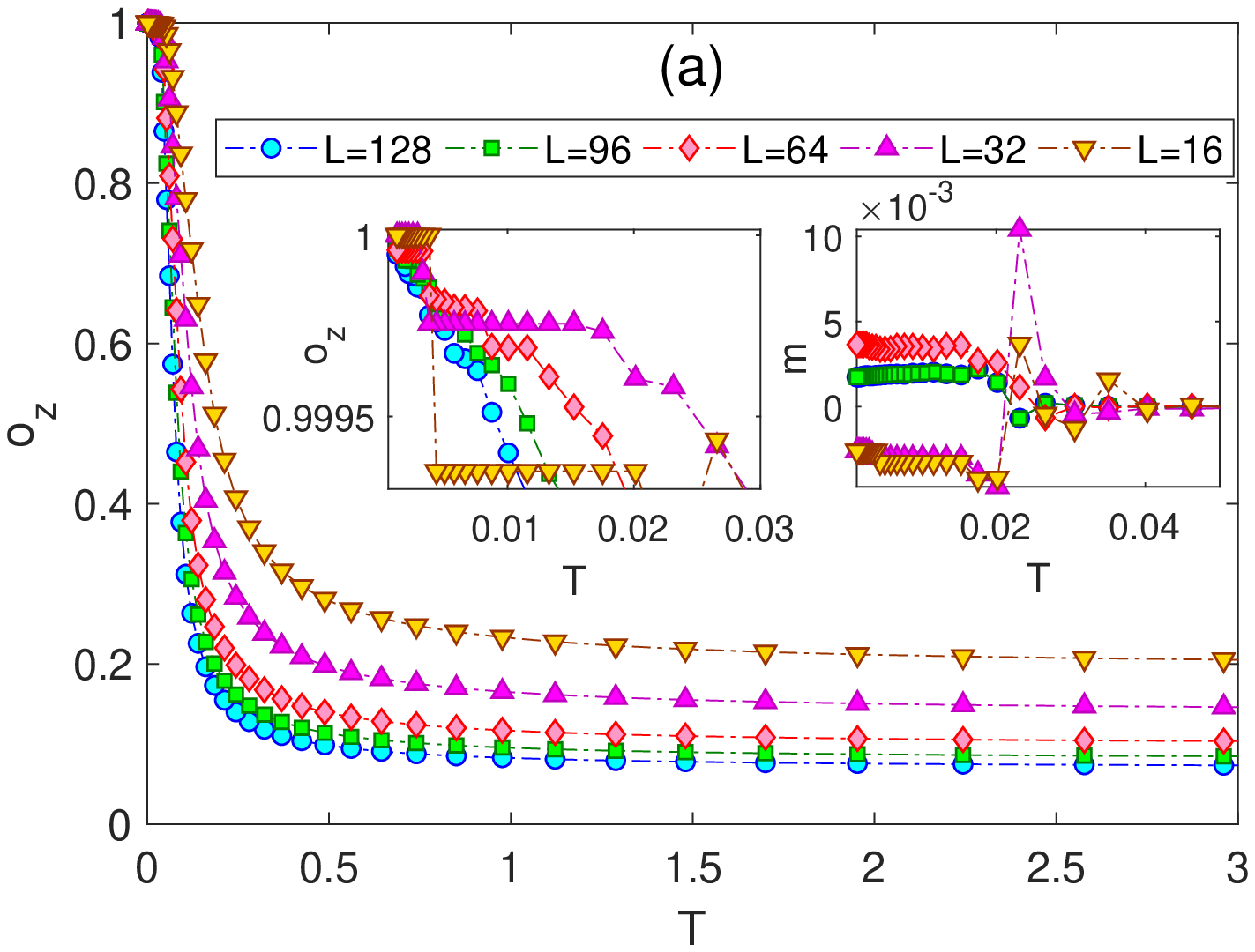}\label{fig:oz_fixedJ2}}
  \subfigure{\includegraphics[scale=0.45,clip]{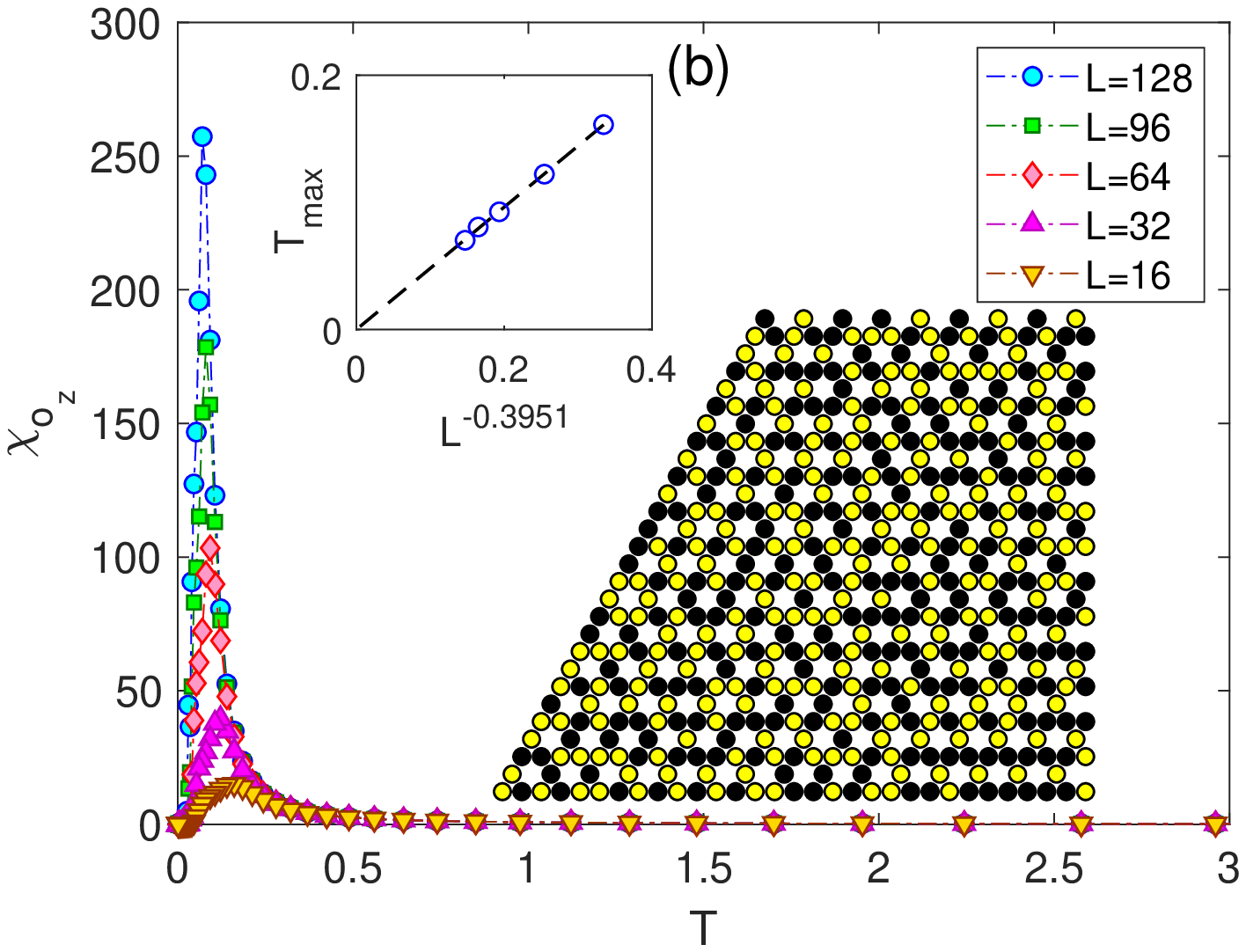}\label{fig:chiz_fixedJ2}}
  \caption{(a) Temperature dependencies of the parameter $o_z$, for $J_2=1/10$ and different lattice sizes $L$. Detailed views of the low-temperature behaviors of $o_z$ and the magnetization $m$ are shown in the insets. (b) The corresponding susceptibility $\chi_{o_z}$ with FSS of its peaks' positions shown in the upper inset. The lower inset shows the snapshot in one representative IAKL layer at the lowest simulated temperature. The black (yellow) circles represent spin-down(up) states.}
  \label{fig:ozfss}
\end{figure}

The chain order parameter $o_z$ is another quantity, which shows a prominent finite-size dependence, as demonstrated in Fig.~\ref{fig:oz_fixedJ2}. With the increasing lattice size the low-temperature increase becomes steeper, which translates into a sharp peak in the corresponding susceptibility $\chi_{o_{z}}$~(Fig.~\ref{fig:chiz_fixedJ2}). With $L \to \infty$ the peaks becomes sharper and higher. However, as the inset demonstrates, their positions converge to $T_{max}(L \to \infty) =
-2.79\times 10^{-05}\pm 0.0043$, i.e., practically zero, which is the critical temperature of a linear chain. The second inset in Fig.~\ref{fig:chiz_fixedJ2} shows a spin snapshot in an arbitrary IAKL layer (all layers show practically the same picture) taken at the lowest simulated temperature with the black and yellow circles representing spin-down and spin-up states, respectively. No apparent LRO is discernible from the snapshot.

At very low temperatures one can observe an interesting ``freezing'' phenomenon, similar to the one reported for the SIATL model~\cite{netz91a,borovsky2016,zukovic2018}. Such a freezing occurs when the chains along the $z$ axis are almost perfectly ferromagnetically ordered but there are still some domain walls separating pieces of the chains, which are ferromagnetically ordered but with spins pointing in opposite directions. At very low temperatures it is difficult to correct such defects, using standard single-spin flip algorithm but, as illustrated by the parameter $o_z$ shown in the first inset of Fig.~\ref{fig:oz_fixedJ2}, after some temperature intervals with frozen intrachain fluctuations eventually all the chains become almost perfectly ordered with the saturated value of $o_z=1$. This freezing is also reflected in the behavior of the total magnetization, shown in the second inset, which also freezes in the corresponding temperature ranges. We note that the very small but finite values of $m$ are only finite-size effects and $m$ is expected to vanish for $L \to \infty$.

\subsubsection{Large interlayer coupling}

\begin{figure}[t!]
  \center
  \subfigure{\includegraphics[scale=0.45,clip]{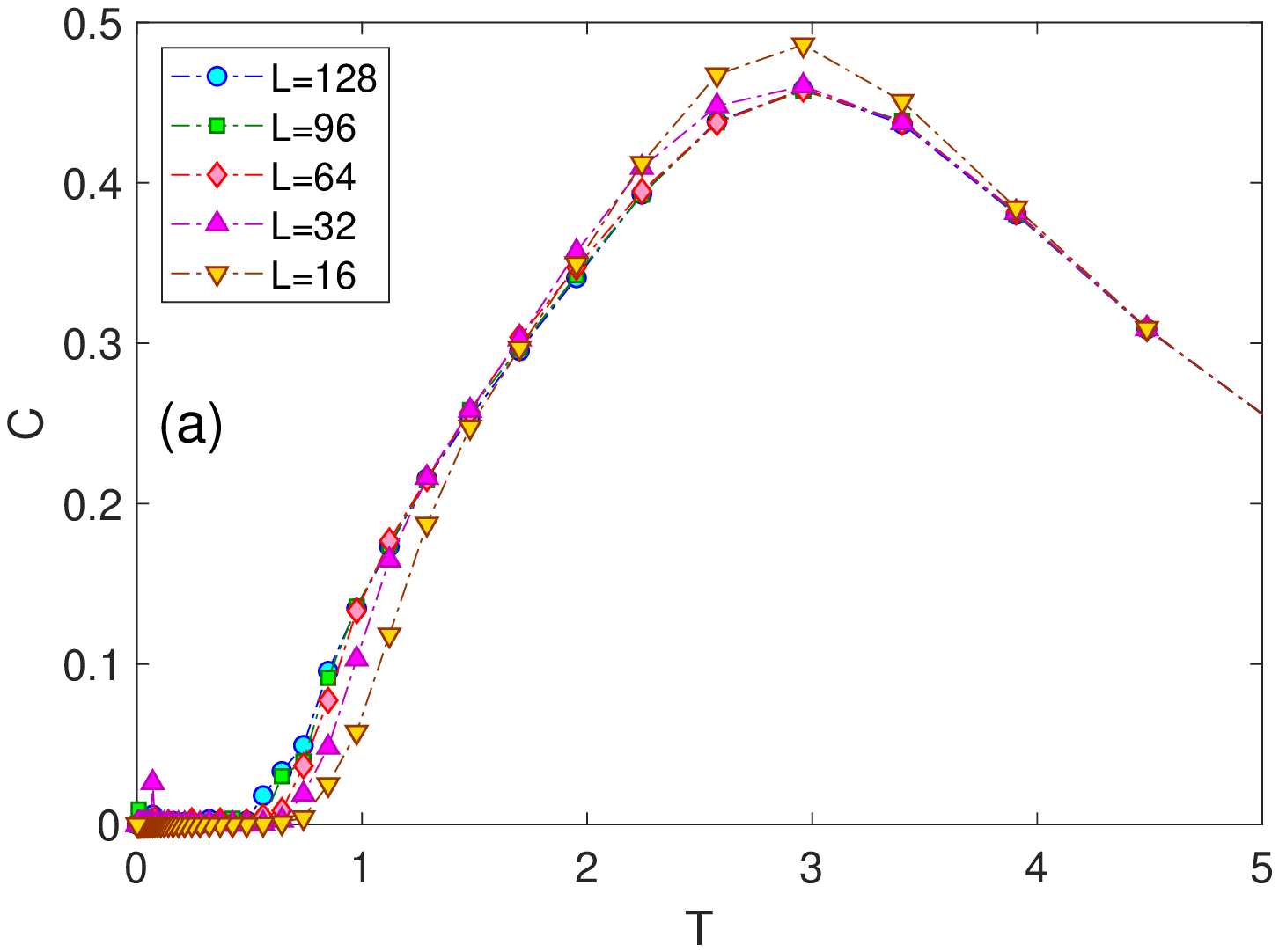}\label{fig:c_fixedJ22}}
  \subfigure{\includegraphics[scale=0.45,clip]{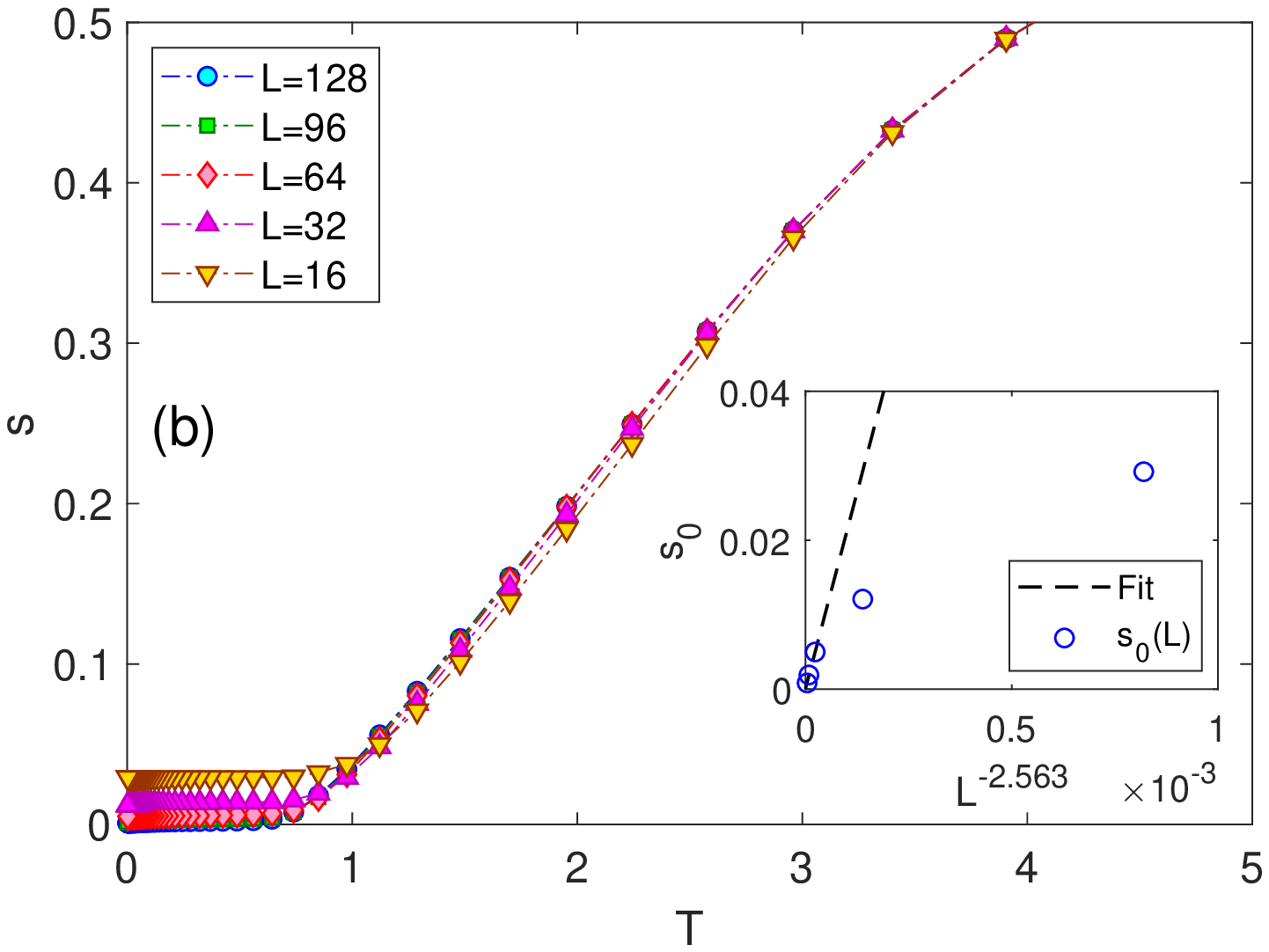}\label{fig:s_fixedJ22}}
	\subfigure{\includegraphics[scale=0.45,clip]{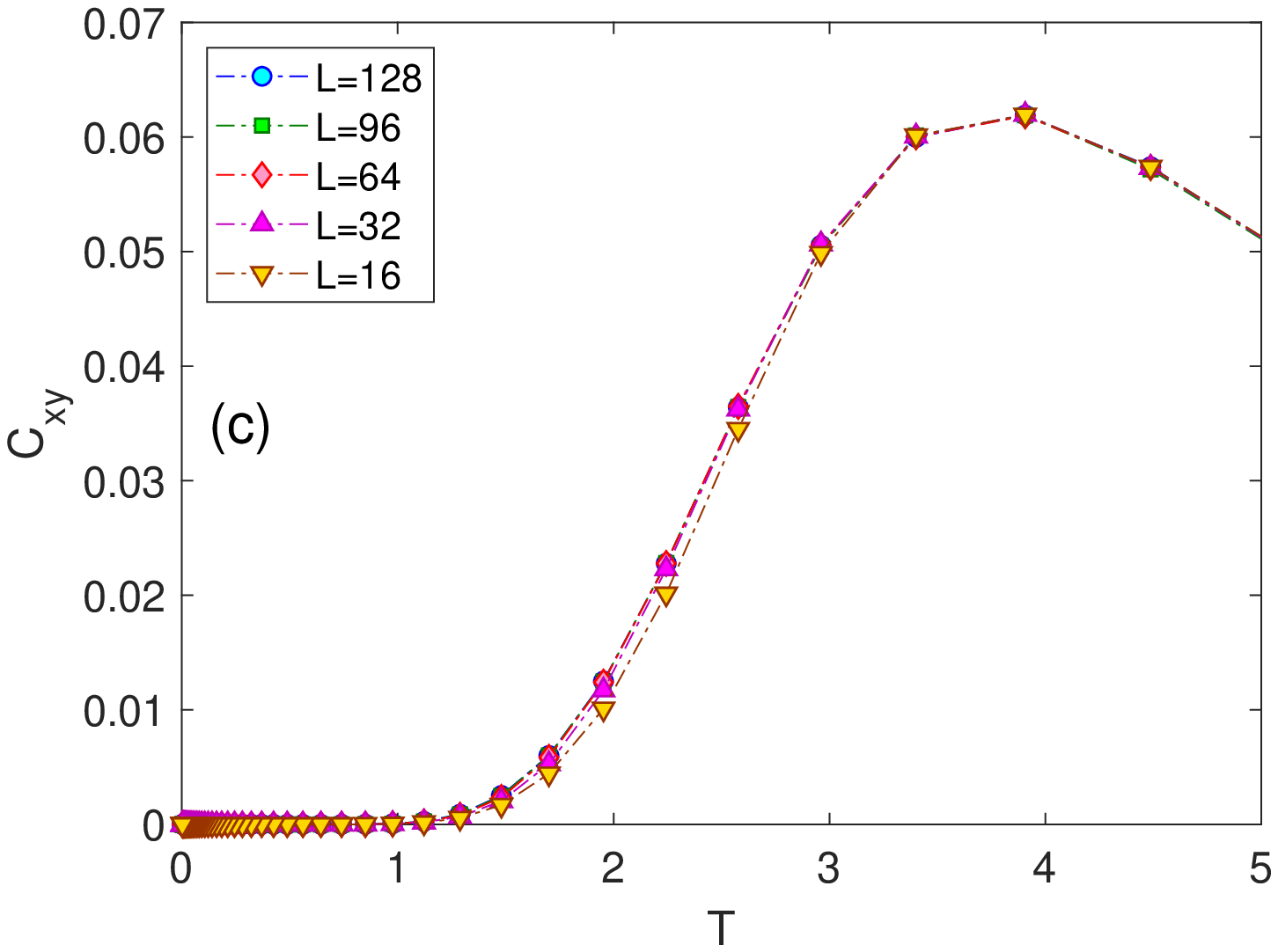}\label{fig:cxy_fixedJ22}}
  \subfigure{\includegraphics[scale=0.45,clip]{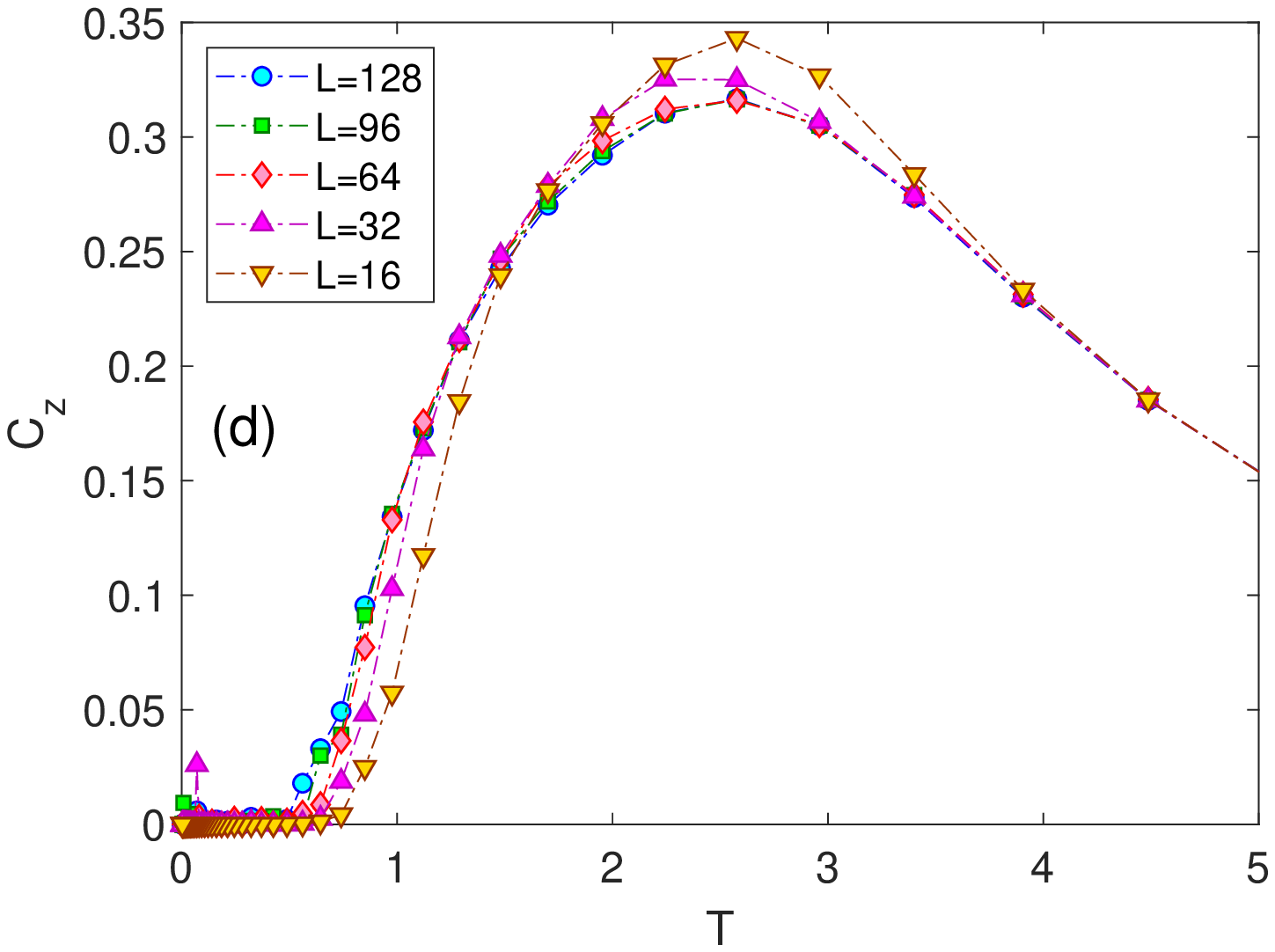}\label{fig:cz_fixedJ22}}
  \caption{(a) The specific heat $C$ and (b) the entropy density $s$, for different values of $L$ and $J_2=2$. The inset in (b) shows the extrapolation of the residual entropy values to thermodynamic limit $s_o(L \to \infty)=6.97\times 10^{-5}\pm
0.0015$, considering only the largest sizes $L$ complying with the asymptotic linear regime. (c) $C_{xy}$  and (d) $C_z$ components of the specific heat $C$.}
  \label{fig:cfss2}
\end{figure}

Even though no signatures of LRO were found for the relatively small interlayer coupling $J_2=1/10$, there is still possibility of LRO at larger $J_2$, as it was the case in the SHAKL model~\cite{gotze16}. Therefore, we also performed a similar FSS analysis for the relatively large $J_2=2$. For such large values the low- and high-temperature peaks of the specific heat are no longer separated but the presence of the subdominant peak can be still seen as a shoulder at the low-temperature side of the dominant broad peak~(Fig.~\ref{fig:c_fixedJ22}). Some lattice-size dependence can be observed in the form of the broad peak decrease at the cost of the shoulder increase towards lower temperatures. This shift can be ascribed to the $C_z$ component (see Figs.~\ref{fig:cxy_fixedJ22} and~\ref{fig:cz_fixedJ22}), which reflects the intrachain fluctuations in the $z$ axis direction. 

The inset in Fig.~\ref{fig:s_fixedJ22} shows the extrapolation of the residual entropy values $s_o(L \to \infty)=6.97\times 10^{-5}\pm 0.0015$, considering only larger sizes ($L \geq 64$) complying with the asymptotic linear regime. Like in the $J_2=1/10$ case, the residual entropy in the thermodynamic limit appears to vanish. In spite of the macroscopic degeneracy, caused by the presence of ``free chains'' similar to the 2D IAKL model, the vanishing residual entropy \emph{density} can be simply explained by the fact that the number of spins in the 3D system increases with $L$ faster than the chain degeneracy.

\begin{figure}[t!]
  \center
  \subfigure{\includegraphics[scale=0.45,clip]{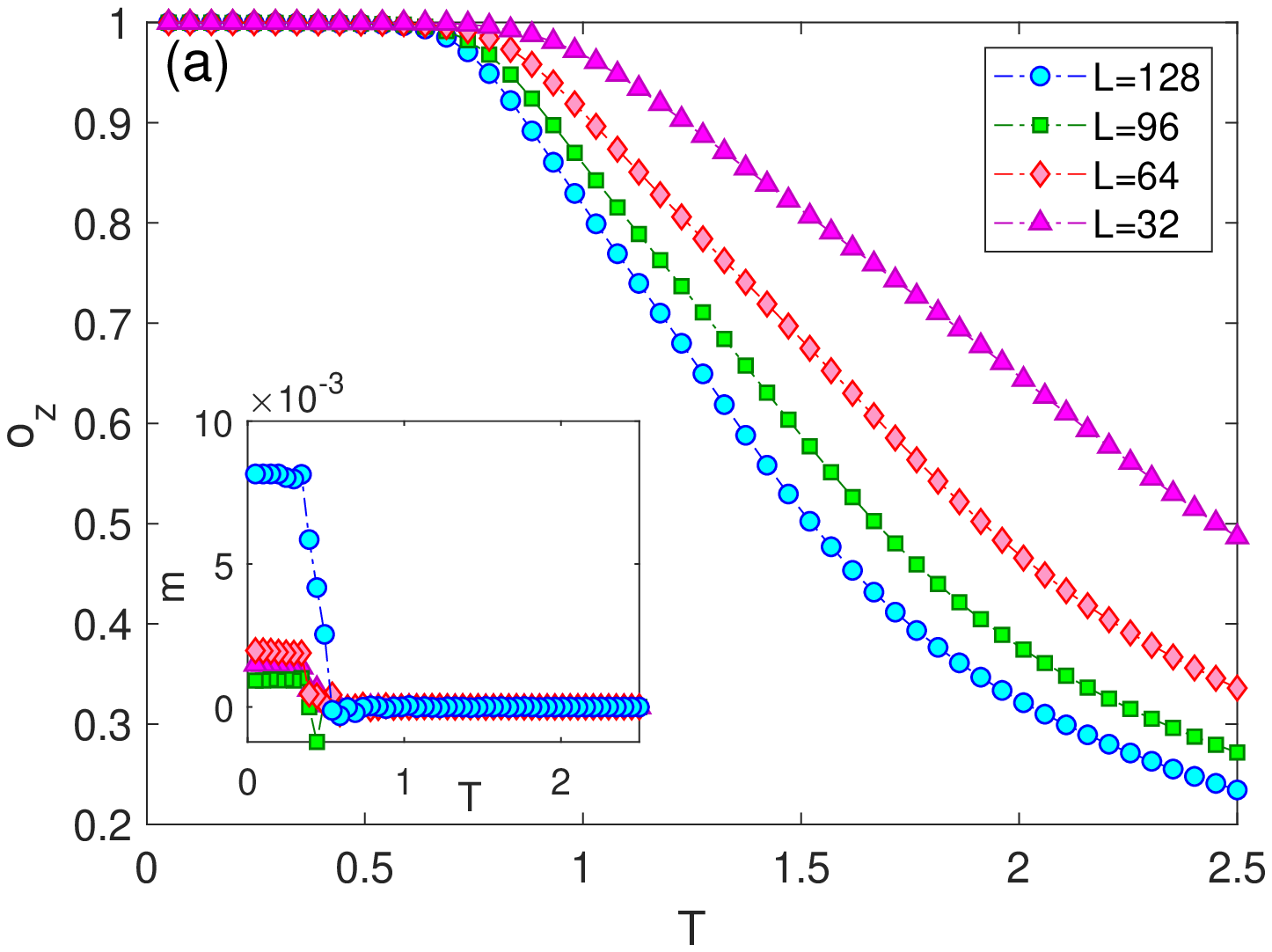}\label{fig:oz_fixedJ22}}
  \subfigure{\includegraphics[scale=0.45,clip]{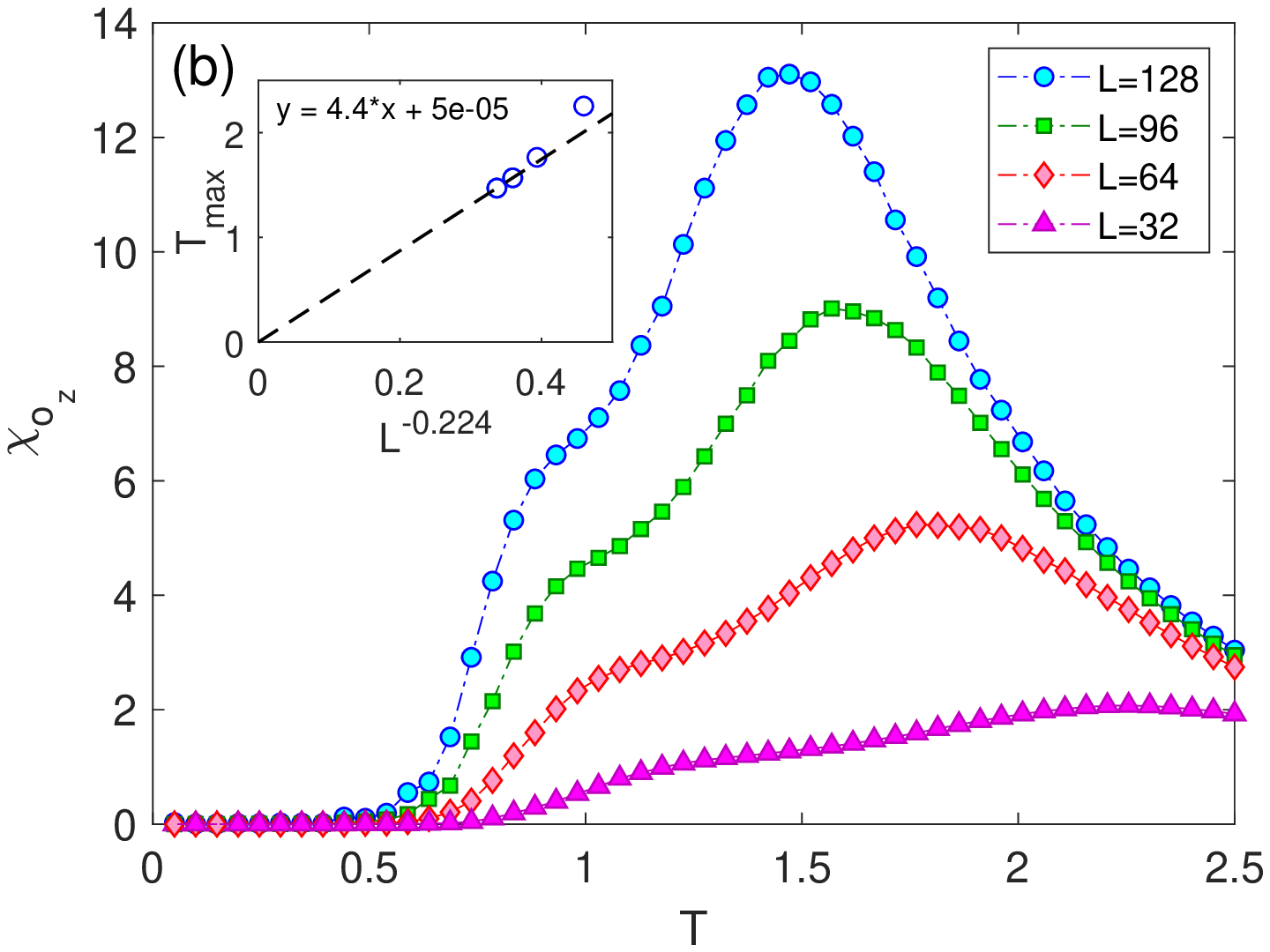}\label{fig:chiz_fixedJ22}}
  \caption{(a) Temperature dependencies of the parameter $o_z$, for $J_2=2$ and different lattice sizes $L$. The inset shows the corresponding total magnetization $m$ curves. (b) The corresponding susceptibility $\chi_{o_z}$ with the finite-size scaling of its peaks' positions shown in the inset.}
  \label{fig:ozfss2}
\end{figure}

Let us now focus on the behavior of the order parameter within the chains in the stacking direction, $o_z$ as well as the sublattice magnetizations, $m_{\alpha}$, which can signal some kind of ordering within the IAKL planes. 
The temperature dependencies of the parameter $o_z$ and its response function $\chi_{o_{z}}$ are presented in Fig.~\ref{fig:ozfss2}. To focus on the anomalies appearing at the intermediate temperatures, we restricted the temperature range to $T \in (0.05,2.5)$ and considered a uniform temperature step. Compared to the $J_2=1/10$ case, the chains show almost perfect ferromagnetic alinement at much higher temperatures, which translates to the shift of the peaks in $\chi_{o_{z}}$ to higher temperatures. Nevertheless, the FSS analysis suggests that, like for $J_2=1/10$, with increasing $L$ these peaks related to the intrachain ordering move to lower temperatures and in the thermodynamic limit $T_{max}(L \to \infty) \to 0$. Additionally, here one can observe shoulders appearing at the low-temperature sides of the peaks (they could not be detected for $J_2=1/10$), which tend to be absorbed by the higher-temperature peaks moving towards $T=0$. The inset in Fig.~\ref{fig:oz_fixedJ22}, showing the total magnetization, indicates that the freezing also occurs below $T \approx 0.5$, albeit it does not show in the behavior of $o_z$ as the chains have already attained almost full ferromagnetic alignment at $T>0.5$.

\begin{figure}[t!]
  \center
  \subfigure{\includegraphics[scale=0.45,clip]{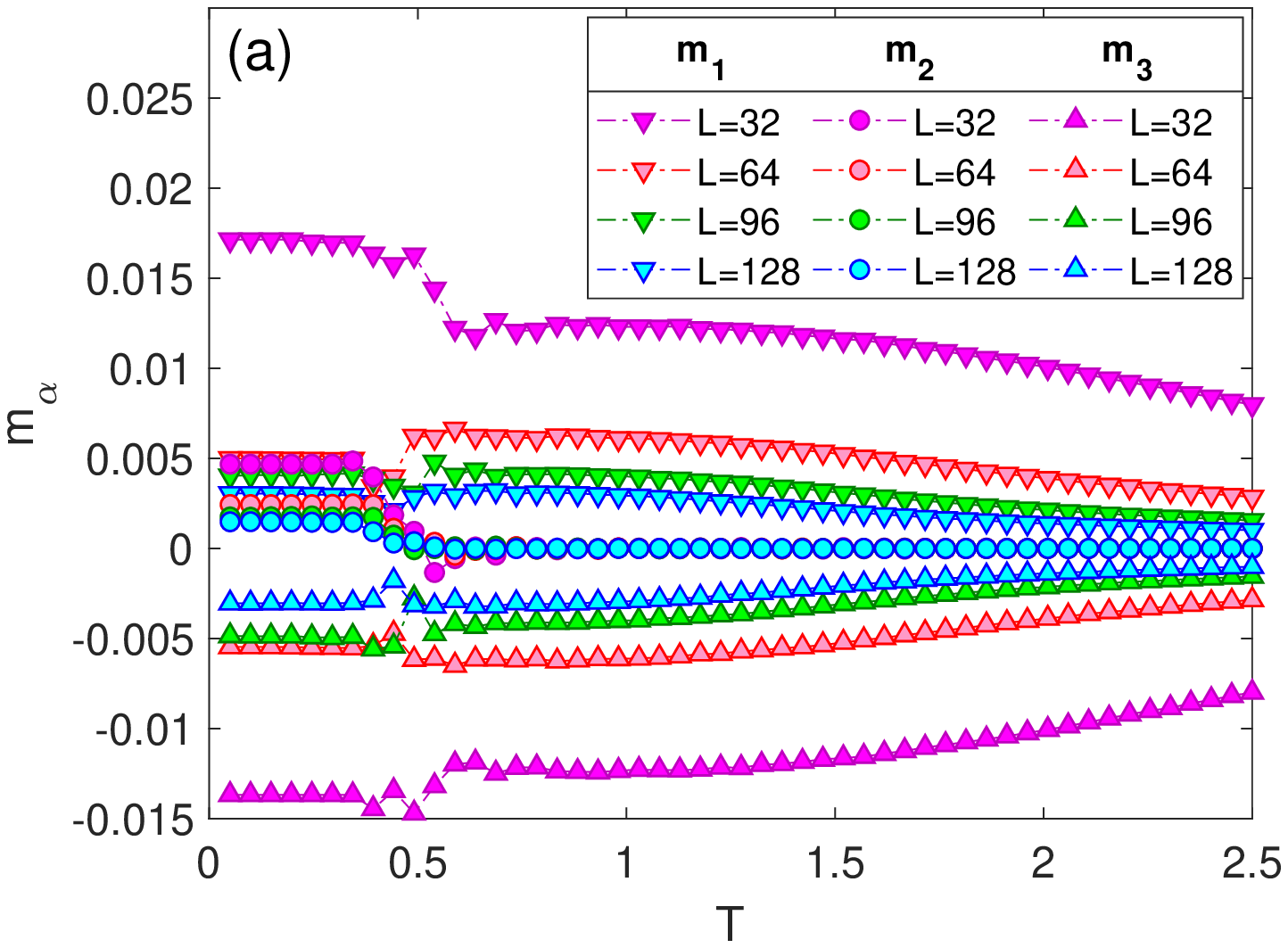}\label{fig:sub_mag_fixedJ22}}
	\subfigure{\includegraphics[scale=0.45,clip]{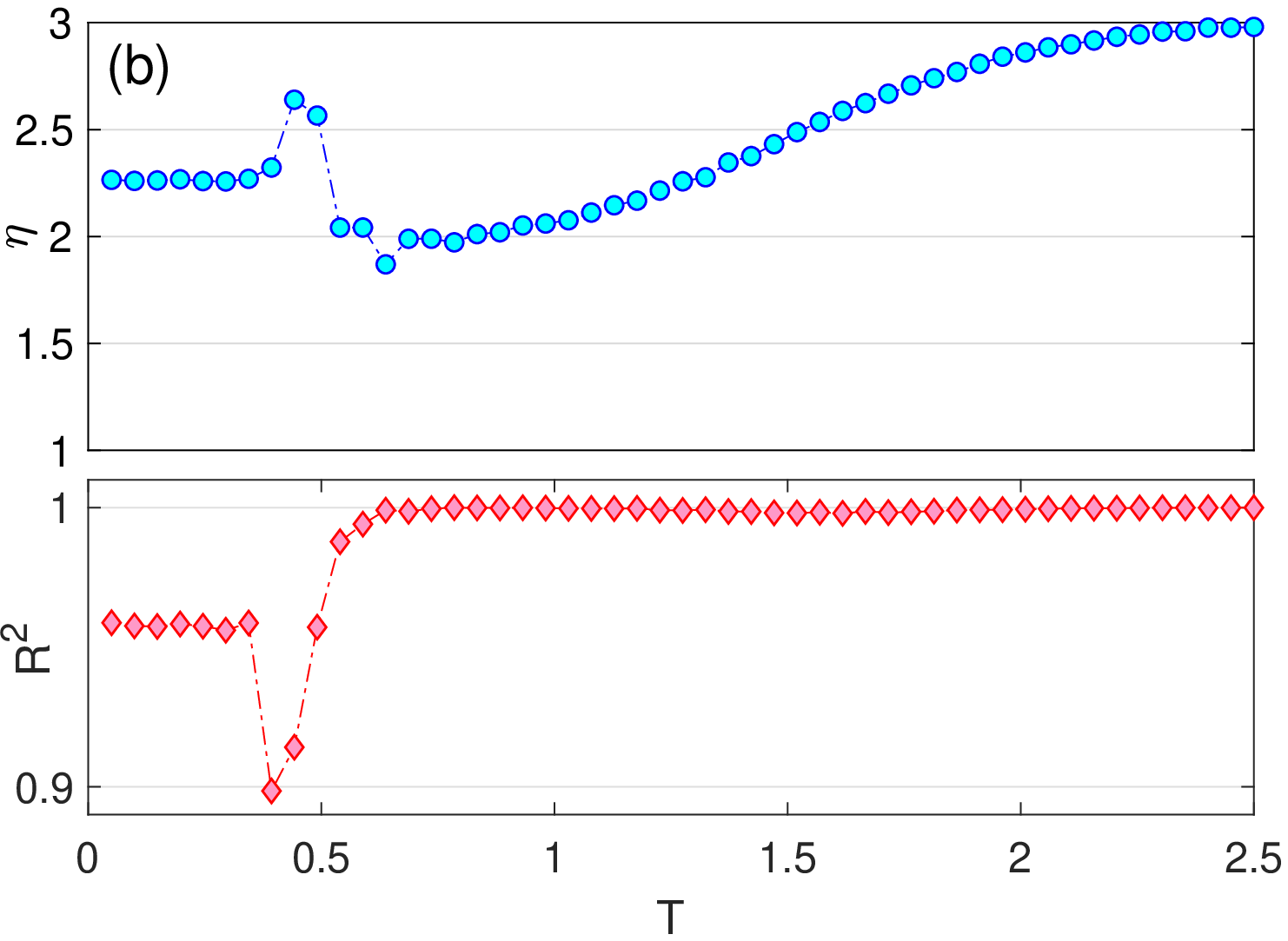}\label{fig:eta_fixedJ22}}
  \caption{(a) Temperature dependencies of the sublattice magnetizations $m_{\alpha}$, for $J_2=2$ and different lattice sizes $L$. (b) The spin-spin correlation function exponent $\eta$ for the order parameter $o=m_1-m_3$ and (c) the corresponding adjusted coefficient of determination $R^2$.}
  \label{fig:sub_mag}
\end{figure}

Finally, we focus on the behavior of the sublattice magnetizations, $m_{\alpha}$, $\alpha=1,2$ and 3, to detect possible LRO within IAKL planes. Let us recall that the related SIATL model displayed the PD LRO with two sublattices ordered antiferromagnetically and the third one disordered. In Fig.~\ref{fig:sub_mag_fixedJ22} the temperature dependencies of the sublattice magnetizations of the SIAKL model are presented for various lattice sizes. For the intermediate temperatures we can see some hints of the PD ordering of the type $(m_1,m_2,m_3)=(M,0,-M)$, like in the SIATL model, however, it does not correspond to any LRO. First of all, the values are very small and with the increasing lattice size tend to vanish. 

To check the possibility of at least algebraic quasi-LRO, like the BKT ordering reported for the SIATL model with a finite number of vertically stacked IATL layers~\cite{lin2014}, we calculated the spin-spin correlation function exponent $\eta$ from the standard FSS relation $o \propto L^{-\eta/2}$, for the potential intralayer order parameter $o=m_1-m_3$. As shown in Fig.~\ref{fig:eta_fixedJ22}, above the freezing temperature $T_f \approx 0.5$ the exponent $\eta$ only takes trivial (system dimensionality related) values corresponding to the paramagnetic phase. It is worth noticing that below $T \approx 1$ the system crosses over to 2D-like paramagnetic behavior with $\eta=2$. In this regime the chains in the stacking direction are ferromagnetically aligned and act like giant Ising spins on 2D lattice. Nevertheless, there is no LRO among the chains. Below $T_f$ the sublattice magnetizations freeze into some nontrivial values, which do not follow the used FSS relation and thus the exponent $\eta$ becomes meaningless. This is also apparent from the adjusted coefficient of determination $R^2$, as a measure of goodness of the linear fit on a log-log scale expected in the algebraic phase~\cite{theil61}, presented in the lower panel of Fig.~\ref{fig:eta_fixedJ22}.


\section{Summary and conclusions}

Motivated by the previously reported emergence of LRO in several highly frustrated 2D systems due to their vertical stacking, 
in the present paper we studied the possibility of such a phenomenon in the Ising antiferromagnet of the kagome lattice (IAKL). It is known that in the 2D IAKL model any LRO is inhibited by a high degree of geometrical frustration. Therefore, we were interested whether the increased dimensionality in the 3D stacked (SIAKL) model can lead to some kind of LRO, as it was the case in the related Ising model on the triangular lattice (IATL) or the Heisenberg model on the present kagome lattice (HAKL). In the former case, first a quasi-LRO and then also true LRO phases appeared by gradual stacking of a finite number of IATL layers~\cite{lin2014}, while the emergence of LRO in the latter required a sufficient strength of the interlayer interaction between the HAKL layers~\cite{gotze16}.  

The results obtained from the present Monte Carlo simulations lead us to the conclusion that no kind of LRO can be established by vertical stacking of either finite or infinite number of the IAKL layers. We looked for some indications of LRO in the SIAKL model with a varying ratio of the interlayer and intralayer coupling strengths $J_2/|J_1| \in [0,2]$. We focused in more detail on two cases of a relatively small $J_2/|J_1|=0.1$ and large $J_2/|J_1|=2$, for which we performed a careful finite-size scaling (FSS) analysis. We found some anomalies, such as round peaks and shoulders in the calculated response functions, nevertheless, those can be associated with non-critical fluctuations within the IAKL planes and the linear chains in the stacking direction. At sufficiently low temperatures the latter order ferromagnetically within each chain, however, there is no LRO among the chains except for the short-range ferrimagnetic arrangement on each elementary triangle, just like individual spins order in the IAKL model. Even the intrachain ordering occurs at finite temperatures only in finite-size systems and in the thermodynamic limit true LRO can only be expected at zero temperature. 

Our findings about no LRO in the present system can also be supported by the arguments provided by the quantum-classical mapping, i.e. using the correspondence between $d$-dimensional quantum systems and $(d+1)$-dimensional classical systems (see, e.g. Ref.~\cite{isakov2003} for the triangular lattice case). In particular, the criticality in our SIAKL classical Ising models as a function of temperature, should be equivalent to the quantum criticality of the 2D Ising model on a kagome lattice in a transverse field. The latter problem has been already investigated by quantum MC and series expansions~\cite{moessner2000,moessner2001,powalski2013}, and it was found that a so-called ``disorder by disorder'' scenario is realized, i.e. there is no quantum phase transition as a function of the transverse field. This finding is fully consistent with our results on the SIAKL model.

At low temperatures we also observed a ``freezing'' phenomenon, similar to the one reported in the SIATL model. We note that unconventional spin freezing and two-dimensional correlation at very low temperatures have been revealed through magnetization and heat capacity measurements in the undistorted Ising kagome antiferromagnet ${\rm MgCo}_3{\rm(OH)}_6{\rm Cl}_2$~\cite{fuji2012}. Theoretical investigations revealed that such a freezing into a spin-glass state in IAKL can be caused by infinitesimal disorder~\cite{schmidt2015} which, however, was not the case in the above experimental study. Therefore, we speculate that the reported freezing behavior might be of the same origin as in the present study and might occur due to the presence of non-vanishing interactions between the kagome planes in the stacking direction.

Finally, our results suggest that SIAKL is a rather rare example of a 3D paramagnet, like the chessboard (CB3) and pyrochlore lattices, but with vanishing zero-point entropy. Another interesting fact is that SIAKL is only frustrated within the layers but there is no frustration in the stacking direction. Such unfrustrated stacking brought about LRO for some range of finite temperatures in the case of triangular but not in the kagome layers. It is also worth mentioning that neither the increased spin magnitude, which led to LRO in the 2D IATL model, was able to stabilize LRO in the 2D IAKL case~\cite{semjan2020}. Therefore, as a future study, we find it interesting to consider a combined effect of the enhanced lattice dimensionality and the increased multiplicity of the local degrees of freedom on the LRO properties of the Ising antiferromagnetic systems with kagome lattice geometry.


\section*{Acknowledgment}
This work was supported by the Scientific Grant Agency of Ministry of Education of Slovak Republic (Grant No. 1/0531/19), the Slovak Research and Development Agency (Grant No. APVV-18-0197), and Internal Scientific Grant System of Faculty of Science of UPJ\v{S} (Grant No. VVGS-PF-2021-1743). The authors would also like to thank the Joint Institute for Nuclear Research in Dubna, Russian Federation, for the use of their Govorun Supercomputer.



\end{document}